\documentclass[prl,twocolumn,showpacs,floatfix,amsmath,amssymb,superscriptaddress]{revtex4}
\usepackage{latexsym}
\usepackage{graphicx}
\usepackage{times,psfrag,subfigure}
\usepackage{amsmath}
\usepackage{dcolumn}
\usepackage{latexsym,amsmath,amssymb,bm,euscript}
\newcommand{\fss}{FeSc$_2$S$_4$}
\newcommand{\mss}{MnSc$_2$S$_4$}

\begin{document}

\title{A Spin-Orbital Singlet and Quantum Critical Point on the Diamond
  Lattice: \fss}\date{\today}

\author{Gang Chen}
\affiliation{Physics Department, University of California, 
Santa Barbara, CA 93106}
\author{Leon Balents}
\affiliation{Kavli Institute for Theoretical Physics, 
University of California, Santa Barbara, CA 93106}
\author{Andreas P. Schnyder}
\affiliation{Kavli Institute for Theoretical Physics, 
University of California, Santa Barbara, CA 93106}

\begin{abstract}
  We present a theory of spin and orbital physics in the A-site spinel
  compound \fss, which experimentally exhibits a broad ``spin-orbital
  liquid'' regime. A spin-orbital Hamiltonian is derived from a
  combination of microscopic consideration and symmetry analysis.  We
  demonstrate a keen competition between spin-orbit interactions, which
  favor formation of a local ``Spin-Orbital Singlet'' (SOS), and
  exchange, which favors magnetic and orbital ordering.
  Separating the SOS from the ordered state is a quantum critical point
  (QCP).  We argue that \fss\, is close to this QCP on the SOS side.
  The full phase diagram of the model includes a
  commensurate-incommensurate transition within the ordered phase.  A
  variety of comparison to and suggestion for experiments are discussed.
\end{abstract}
\date{\today}

\pacs{71.70.Ej,71.70.Gm,75.10.-b,75.40.-s}

%\email{chggst@physics.ucsb.edu}

\maketitle

%%%%%%%%%%%%%%%%%%%%%%%%%%%%%%%%%%%%%%%%%%%%%%%%%%%%%%%%%%%%%%%%%%%%%%%%%%%%%
The search for quantum spin liquids -- materials in which local moments
are well formed but continue to fluctuate quantum mechanically down to
zero temperature -- is a fundamental challenge in condensed matter
physics.  In transition metal compounds with relatively high
(e.g. cubic) symmetry, a richer possibility has also been contemplated,
in which not only spin but also {\sl orbital} states of localized
electrons fluctuate.  Such a ``Spin Orbital Liquid'' (SOL) was proposed
in LaTiO$_3$ \cite{khaliullin2000olt}.  Probably the best
candidate for a SOL is \fss\, a spinel compound (with the general
structure AB$_2$X$_4$), in which only the A sites form a
magnetically/orbitally active {\sl diamond} sublattice.  In recent
years, a variety of such A-site spinels, e.g. CoAl$_2$O$_4$ and \mss,
were also found to be
frustrated \cite{giri:prb,krimmel:prb1,tristan:prb05,krimmel:pb,suzuki:jpcm},
forming a ``spiral spin liquid'' \cite{bergman:natp} at certain
temperature range. \fss\ stands out markedly amongst this class of
compounds in exhibiting a much broader liquid regime, extending down to
the lowest measured temperatures.

\begin{figure}[hbtp]
\includegraphics[width=8.0cm]{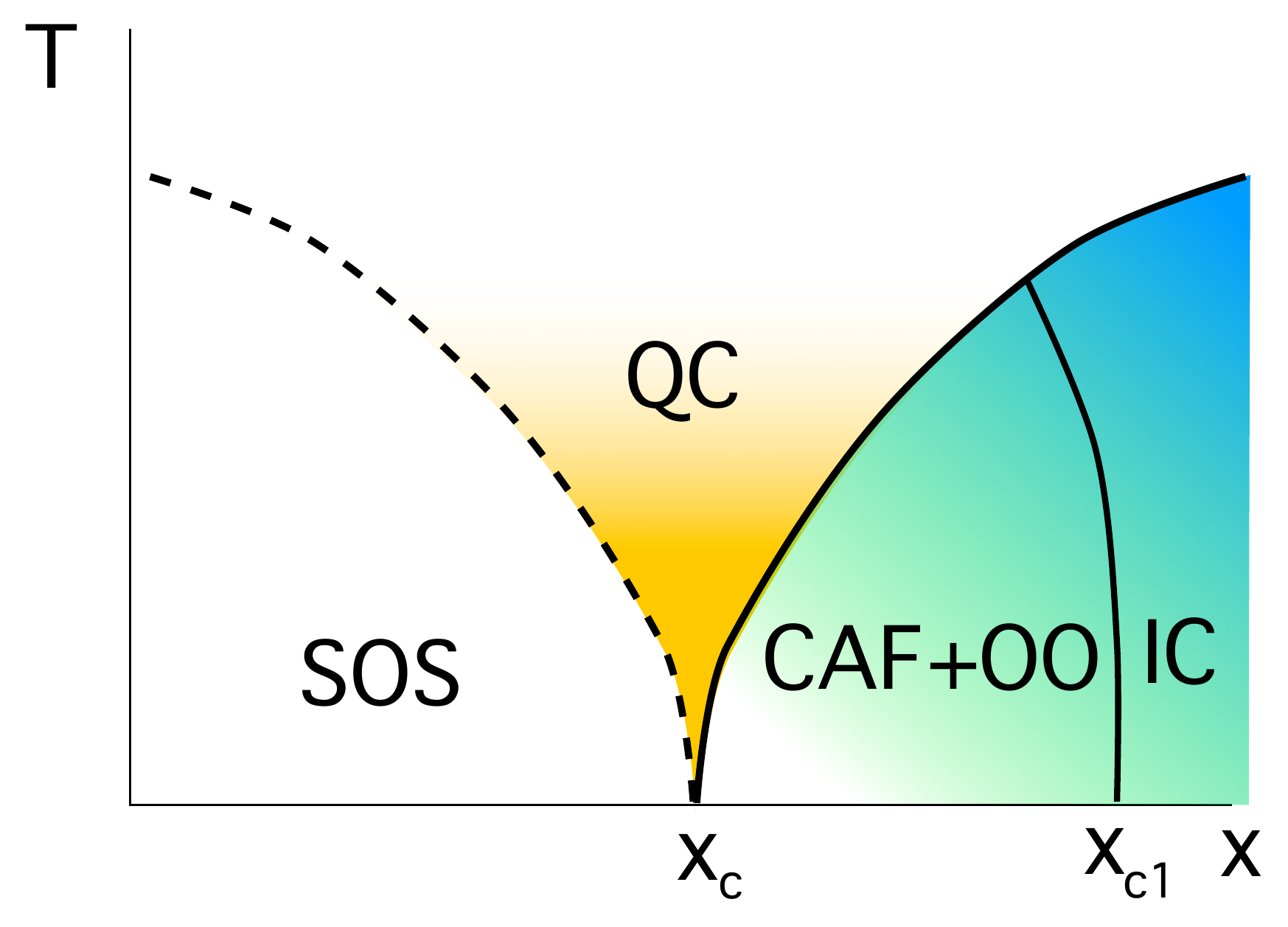}
\caption{(color online).  The schematic phase diagram versus $T$,
  temperature and $x$, the ratio of exchange to spin orbit interaction
  (see text).  Here the labels for the phases are: ``SOS'',
  spin-orbital-singlet; ``CAF+OO'', commensurate antiferromagnet with
  orbital order; ``IC'', incommensurate spin and orbital order.  ``QC''
  stands for the quantum critical regime.}
\label{fig:pd}
\end{figure}
In the letter we describe a theoretical study of the spin and orbital
physics of \fss\,
\cite{loidl:nmr,loidl:ns,loidl:nsnmr,fichtl:05,fritsch:prl04}. Here, the
magnetic Fe$^{2+}$ ion is in a $3d^6$ configuration, with a local $S=2$
moment as well as a two-fold {\sl orbital degeneracy} (associated with
one hole in the $e_g$ doublet), i.e.  Fe$^{2+}$ is Jahn-Teller
active. However, no indication of long range spin or orbital order is
detected down to $50$mK, which makes the frustration parameter $f\gtrsim
1000$ ($f \equiv \Theta_{CW}/T_N$ and $\Theta_{CW}=-45.1\text{K}$ is the
Curie-Weiss temperature), one of the largest values ever observed.  We
argue that the unusual behavior of \fss\ arises from proximity to a
Quantum Critical Point (QCP) between a spin-orbital singlet phase
induced by atomic spin-orbit coupling and a magnetically and orbitally
ordered phase favored by exchange (see Fig.~\ref{fig:pd}).

%%%%%%%%%%%%%%%%%%%%%%%%%%%%%%%%%%%%%%%%%%%%%%%%%%%%%%%%%%%%%%%%%%%%%%%%%%%%%%
\emph{The spin-orbital Hamiltonian in \fss} --- The na\"ively large
(10-fold) spin and orbital degeneracy of the isolated Fe$^{2+}$ ions
must be split by additional effects.  One natural mechanism is exchange.
To study this problem, we will first derive a ``Kugel-Khomskii'' type
spin-orbital exchange Hamiltonian \cite{kk:82}, in which the spin and
orbital state on each Fe$^{2+}$ are described by spin ${\bf S}_i$
($S=2$) and pseudo-spin $\boldsymbol{\tau}_i$ ($\tau=1/2$).  The study
of a structurally identical A-site spinel compound \mss\,(no orbital
degeneracy) indicates \cite{bergman:natp,kim:prl,sungbin:08} that a
minimal superexchange Hamiltonian should include both the
nearest-neighbor (NN) and next-nearest-neighbor (NNN)
interactions. Unfortunately, direct derivation of the superexchange
Hamiltonian from the ``parent'' Hubbard model is not very practical due
to the complicated exchange paths A-X-B-X-A \cite{roth:64} which involve
{\sl five ions} for both NNs and NNNs. However, the Fd$\overline{3}$m
space group symmetry of \fss\, rather strongly constrains the exchange
Hamiltonian.  One can show that the most general spin-orbital exchange
Hamiltonian allowed by symmetry for NNs and NNNs has the following simple
form (neglecting the spin anisotropic terms),
\begin{eqnarray}
{\mathcal H}_{ex} 
&=& \tfrac{1}{2}\sum_{ij} [J_{ij} {\bf S}_i \cdot {\bf S}_j 
+ 8K_{ij} \boldsymbol{\tau}_i \cdot \boldsymbol{\tau}_j 
+ \tilde{K}_{ij} \boldsymbol{\tau}_i^y \boldsymbol{\tau}_j^y 
\nonumber \\
& & + (L_{ij}  \boldsymbol{\tau}_i \cdot \boldsymbol{\tau}_j +
\tilde{L}_{ij} \boldsymbol{\tau}_i^y \boldsymbol{\tau}_j^y)
{\bf S}_i \cdot {\bf S}_j]
\;,
\label{eq:syham}
\end{eqnarray}
where $J_{ij}=J_1$ or $J_2$ when $ij$ are NN and NNN sites, respectively
(and similarly for $K_{ij},\tilde{K}_{ij},L_{ij},\tilde{L}_{ij}$).

To further constrain the couplings, we must treat the microscopic
physics.  Integrating out the intermediate states on B and X sites, we
consider processes involving electron motion between two Fe$^{2+}$ ions.
In general, superexchange may occur between electrons from either the
$t_{2g}$ or $e_g$ levels.  Only processes involving the latter, however,
distinguish the orbital pseudo-spin.  Now note that in
Eq.~(\ref{eq:syham}), there are no terms linear in the pseudo-spin
operators $\boldsymbol{\tau}$.  Therefore, microscopic processes which
might individually contribute to such terms for a single pair of
Fe$^{2+}$ ions must cancel when summed together.  Orbital contributions
from superexchange processes {\sl between} $e_g$ and $t_{2g}$ are
precisely of this form, and can therefore be neglected.  Thus to obtain
the orbital part of the exchange Hamitonian, we only need to focus on
terms involving motion of two $e_g$ electrons.  Again, symmetry
restricts the form of Hamiltonian describing the effective transfer
between NNs and NNNs:
\begin{eqnarray} 
  {\mathcal H}_t &=& \sum_{ij} \sum_{m\sigma} t_{ij}
  d_{im\sigma}^{\dagger}d_{jm\sigma}^{\vphantom\dagger}
  \;, 
\end{eqnarray} 
where $d_{im\sigma}^{\dagger}$ and $d_{im\sigma}$ are the creation
and annihilation operators of a hole at $m$th $e_g$ level with spin
$\sigma$, and $t_{ij}=t_1,t_2$ when $i,j$ are NNs and NNNs,
respectively. Together with the on-site Hubbard-U terms, one can derive
the pseudo-spin part of the exchange Hamiltonian by the standard
perturbation theory. Combined with the contribution to the spin part
from the $t_{2g}$-$t_{2g}$ and $t_{2g}$-$e_g$ exchange, we obtain a
simplified version of Eq.~(\ref{eq:syham}) with
\begin{eqnarray}
  \label{eq:1}
  \tilde{K}_{ij}=\tilde{L}_{ij}=0, \qquad
  L_{ij} = 2K_{ij} ,
\end{eqnarray}
with $K_{1,2} =  t_{1,2}^2/(4U) > 0$.  The simplified Hamiltonian contains
4 dimensionful parameters ($J_1,J_2,K_1,K_2$).  Moreover, from the above
analysis, we expect crudely $J_{1,2} \gg K_{1,2}$ as the $K$'s only
come from the $e_g$-$e_g$ exchange.

The exchange Hamiltonian ${\mathcal H}_{ex}$ has spin-rotational
symmetry.  A second mechanism to split the large ionic degeneracy is
spin-orbit coupling.  For an isolated Fe$^{2+}$ ion, the physical spin
and pseudo-spin can couple via the symmetry allowed term \cite{vallin:prb},
\begin{equation}
  {\mathcal H}_0^i = -\frac{\lambda}{3} \{ \sqrt{3}
  \tau_i^x [(S_i^x)^2-(S_i^y)^2] + \tau_i^z [3(S_i^z)^2-{\bf S}_i^2] \}
  \;.
  \label{eq:onsiteH}
\end{equation}
Note that, for an isolated single Fe$^{2+}$ ion, ${\mathcal H}_0^i$
results in a {\sl non-degenerate} ground state with a gap $\lambda$ to
the first excited triplet.  The ground state is a highly entangled state
of spin and orbital degrees of freedom: a ``spin orbital singlet''.   It
is remarkable that a 3d ion with large spin $S=2$ can form such a highly
entangled quantum ground state.

Such spin orbital singlet formation competes with exchange, so it is
helpful to have a microscopic estimate of $\lambda$.  This interaction
arises at {\sl second} order in the $LS$ spin-orbit interaction
$\lambda_0 ({\bf L}\cdot{\bf S})$ due to the nonvanishing matrix
elements of ${\bf L}$ between $e_g$ and $t_{2g}$ orbitals.  Standard
second order perturbation theory gives \cite{vallin:prb} $\lambda
=6 {\lambda_0}^2/\Delta_{te} >0$, with $\Delta_{te}$ the crystal
field splitting between $e_g$ and $t_{2g}$ levels. Taking the atomic
spin-orbit coupling constant $|\lambda_0| \approx 100 \text{cm}^{-1}$ and
$\Delta_{te} \approx 2500 \text{cm}^{-1}$ \cite{testelin:prb,feiner:jpc}
yields $\lambda \approx 36 \text{K}$.  It is noteworthy that this is
comparable to $\Theta_{CW}$.  If we assume $K_1=K_2=0$, by the high temperature mean field
theory (including both $\sum_i{\mathcal H}_0^i$ and ${\mathcal H}_{ex}$)
one finds $\Theta_{CW} =-\tfrac{S(S+1)}{3} (4J_1+12J_2)$.  
Thus $\Theta_{CW}$ is a measure of the strength of exchange, and we
conclude that exchange and spin-orbit coupling are competitive in \fss.

% Thus, from the numerical values, we should have $\lambda \gg J_{1,2}$.
% The on-site spin-orbital interaction ${\mathcal H}_0^i$ has a non-degenerate ground state,
% and $\langle{\bf S}_i\rangle  = \langle \boldsymbol{\tau}_i \rangle = \langle {\bf S}_i \boldsymbol{\tau}_i \rangle = 0$ 
% for this ground state. Furthermore, the spin susceptibility
% $\chi_i(T)$ indeed saturates to a constant $\frac{4(g\mu_B)^2}{\lambda}$ as $T \rightarrow 0$.

From the above analysis, our complete Hamiltonian for \fss\ is, in the
first approximation
\begin{equation}
{\mathcal H} = \sum_i{\mathcal H}_0^i + {\mathcal H}_{ex} 
\;.
\label{eq:fullH}
\end{equation}

%%%%%%%%%%%%%%%%%%%%%%%%%%%%%%%%%%%%%%%%%%%%%%%%%%%%%%%%%%%%%%%%%%%%%%%%%%%%%%%%%%%%%%%%%%
\emph{Minimal model for FeSc$_2$S$_4$} ---
We begin the analysis by considering a simplified exchange Hamiltonian
by appealing to the neutron scattering measurements \cite{loidl:ns},
which indicate the low energy magnetic excitations are localized near
${\bf k}=2\pi(1,0,0)$.  This is precisely the ordering wavevector
associated with a simple collinear N\'eel state on an FCC sublattice,
and suggests the dominance of second neighbor exchange $J_2$.  Therefore
we begin by studying the ``minimal model'' with $J_1=K_1=K_2=0$ and $J_2
>0$ antiferromagnetic.  The two FCC sublattices of the diamond lattice
decouple in this case, and $J_2$ can be viewed as a nearest-neighbor
antiferromagnetic exchange within either of these sublattices.
(Pathological effects of this decoupling can be removed by including
very small $J_1$).

Here there is a single dimensionless parameter
$x \equiv J_2/\lambda$.  For $x\gg 1$, the exchange is dominant, and
since the $S=2$ spins are rather classical, we expect them to order
magnetically.  The ground states of the NN FCC sublattice are
well-known.  In real space, they consist of simple
collinear antiferromagnetic N\'eel states within each
$\{\textrm{100}\}$ plane, with an arbitrary choice of axis for
each such plane.  In momentum space, this allows for spiral states with
wavevector ${\bf k} = 2\pi(1,\delta,0)$ (and symmetry-related
wavevectors) with arbitrary $\delta$.  

The $\lambda$ term splits this degeneracy.  For an arbitary magnetically
ordered state, in which $\langle {\bf S}_i\rangle\neq 0$, the spin
orbital Hamiltonian Eq.~(\ref{eq:onsiteH}) induces an ``orbital field''
that induces an orbital moment on each site.  The magnitude of this
orbital field is maximal when the spin is along one of the axial
directions $[\textrm{100}]$.  This selects {\sl
  commensurate} states with $\delta=0,1/2$.  Within the minimal model,
the remaining degeneracy is lifted by the weak effects of
quantum fluctuations {\cite{chen:progress}}, which favors collinear spin states. 
Note that this selects $\delta=0$, which corresponds to the experimentally 
observed low energy excitations in \fss.  

In contrast to the commensurate ordered phase (with collinear orbital
order) found for $x\gg 1$, for $x\ll 1$, the ground state is a spin
orbital singlet, with a gap to all excited states.  This is a
generalization of a ``quantum paramagnet'' discussed intensively in
spin-only models.  The gap decreases steadily upon reducing $x$, and is
expected to close at a Quantum Critical Point (QCP).

Indeed, this expectation is confirmed by a simple mean field theory
(MFT).  This consists of decoupling the exchange term as usual into an
effective Zeeman field which is self-consistently determined for each
site. Note that this procedure involves no approximation for ${\mathcal
  H}_0^i$.  Assuming an ordered state of the form of a coplanar spiral
\begin{equation}
\langle {\bf S}_i \rangle = m [\cos{({\bf p}\cdot {\bf r}_i)}\hat{x} + \sin{({\bf p}\cdot {\bf r}_i)}\hat{y}],
\label{eq:mft1}
\end{equation}
with ${\bf p}= 2\pi(1,0,0)$ or ${\bf p}= 2\pi(1,1/2,0)$, we predict by
mean field theory that at $T=0$ a continuous second order transition
occurs at $x_c=1/16$, and the staggered magnetization vanishes for $x
\gtrsim x_c$ like
\begin{equation}
m \sim 8\sqrt{2(x-x_c)}
\;.
\label{eq:mft2}
\end{equation}

In the vicinity of the QCP, one may obtain a Landau expansion of the
effective action by standard methods \cite{sachdev1999qpt}. The order
parameters are the (real) staggered magnetizations ${\boldsymbol
  \Phi}_a$ at wavevectors $2\pi
{\bf\hat x}, 2\pi{\bf\hat y},2\pi{\bf\hat z}$ (for $a=1,2,3$):
\begin{equation}
  \label{eq:2}
  \langle {\bf S}_i \rangle = \sum_{a=1,2,3} {\boldsymbol \Phi}_a
  (-1)^{2 x_i^a} ,
\end{equation}
where the $x_i^a$ are the usual half-integer coordinates of the FCC
sites using a unit length conventional cubic unit cell.  The
symmetry-allowed form of the effective lagrangian (in imaginary time $\tau$)
is
\begin{eqnarray}
  \label{eq:3}
  {\mathcal L} & = & \sum_a \Big[|\partial_\tau {\boldsymbol \Phi}_a|^2 + v^2
  |{\boldsymbol\nabla\Phi}_a|^2 + r |{\boldsymbol\Phi}_a|^2 \Big]
  \nonumber \\
  & & + {\rm Sym}\big[ g_1(\Phi_1^x)^4 + g_2 (\Phi_1^x)^2(\Phi_1^y)^2 +
  g_3 (\Phi_1^x)^2(\Phi_2^x)^2 \nonumber
  \\
  && + g_4 (\Phi_1^x)^2(\Phi_2^y)^2 + g_5 \Phi_1^x \Phi_1^y \Phi_2^x \Phi_2^y \Big],
\end{eqnarray}
where ``${\rm Sym}$'' indicates symmetrization with respect to both
wavevector (lower) and spin (upper) indices, and we have for simplicity
neglected presumably unimportant anisotropy of the gradient terms.
Note that the effective Hamiltonian in Eq.~(\ref{eq:fullH}) has actually
{\sl independent} cubic ``internal'' spin symmetry and cubic ``external''
space group symmetry, which both constrain Eq.~(\ref{eq:3}).

This is an Euclidean multi-component $\Phi^4$ field theory of standard
type, which in $D=d+1=4$ space-time dimensions is in its upper critical
dimension.  Thus MFT is expected to be qualitatively correct, up to
logarithmic corrections.  Numerous properties of the ideal QCP follow
directly.  The gap $\Delta$ vanishes upon approaching the QCP from the
spin orbital singlet phase, according to $\Delta \sim \sqrt{x_c-x}$.
Similarly, the N\'eel temperature vanishes approaching from the other
side, $T_N \sim \sqrt{x-x_c}$.  Other critical properties are readily
obtained from the theory of a free relativistic scalar field, up to
logarithmic corrections. A comparison with known experimental results is
given at the end of this letter.

Having established the fundamental nature of the phase diagram and QCP,
we turn to a discussion of more subtle effects.   

%%%%%%%%%%%%%%%%%%%%%%%%%%%%%%%%%%%%%%%%%%%%%%%%%%%%%%%%%%%%%%%%%%%%%%%%%%%%%%%%%%%%%%%%%%
\emph{Commensurate to incommensurate transition in the ordered phase}
---  We first consider the effects of exchanges other than $J_2$ in the
ordered phase.   Define now $x \equiv {\rm max} \{J,K\}/\lambda$, 
where $\{J,K\}$ denotes all exchange coupling constants $J_1,J_2$ and
$K_1,K_2$.  
 In the extreme limit $x\gg 1$, in which the spin orbit interaction
can be neglected, one expects incommensurate magnetically and orbitally
ordered ground states.  If we assume the spins (pseudo-spins) form a coplanar
spiral with wavevector ${\bf p}$ (${\bf q}$) and phase shift $\theta$
($\phi$) between the two fcc sublattices, we obtain 8 conditions for
such a configuration to be a classical ground state:
\begin{eqnarray}
\left\{
\begin{array}{ll}
|\Lambda({\bf p})| = \frac{|J_1|}{8J_2},
&\theta= arg(\Lambda({\bf p}));
\\
|\Lambda({\bf q})| = \frac{K_1}{8K_2}, 
&\phi= arg(\Lambda({\bf q}));
\\
|\Lambda({\bf p} + {\bf q})| = \frac{K_1}{8K_2},
&\theta+\phi= arg(\Lambda({\bf p+\bf q}));
 \\
|\Lambda({\bf p} - {\bf q})| = \frac{K_1}{8K_2},
&\theta-\phi= arg(\Lambda({\bf p-\bf q})).
  \end{array} \right.
\label{eq:mcond}
\end{eqnarray}
Here the complex function $\Lambda({\bf p})$ is defined as
\begin{eqnarray}
\Lambda({\bf p}) = 
 \cos{\frac{p_x}{4}} \cos{\frac{p_y}{4}} \cos{\frac{p_z}{4}} 
                  - i \sin{\frac{p_x}{4}} \sin{\frac{p_y}{4}} \sin{\frac{p_z}{4}}.
\end{eqnarray}
In Eq.~\eqref{eq:mcond}, after  trivially eliminating $\phi,\theta$,
there are actually {\sl six independent equations} which can completely
determine the six real components of ${\bf p}$ and ${\bf q}$.  When some
coupling constants vanish, a variety of different
degenerate classical ground states can be obtained \cite{chen:progress}.
For example, when $K_1=K_2=0$, ${\mathcal H}_{ex}$ reduces to the
familiar $J_1$-$J_2$ model and the degenerate spin spiral momenta ${\bf
  p}$'s form the well-known spiral surface in momentum space \cite{bergman:natp,sungbin:08,kim:prl}.

When we turn on the spin-orbital interaction ($\lambda \neq 0$),
we expect the spin and orbital ordering to become commensurate with
increasing $x$.  This is because, as remarked above, ${\mathcal H}_0^i$
has axial cubic anisotropy, and moreover, it favors orbital order with
twice the momentum of the spin spiral.  A general spin spiral for which
all spins are axially oriented satisfies ${\bf p} =
\frac{\pi}{2}(n_1,n_2,n_3)$, with $n_1,n_2$ and $n_3$ either all even
integers or all odd integers.  Assuming $J_1/(8J_2)$ is not too large
(as expected both from comparison with the structurally similar material
\mss, and from the aforementioned neutron data), the states with ${\bf
  p}= 2\pi(1,0,0),2\pi(1,\tfrac{1}{2},0)$ have low exchange energy, and
are favored by the gain from the spin orbit interaction. Therefore we
expect the ordered state to become commensurate for $x_c < x < x_{c1}$
(see Fig.~\ref{fig:pd}). For the $J_1$--$J_2$--$\lambda$ model, by
comparing the classical energies given by the incommensurate spiral
momenta from the spiral surface $\Lambda({\bf p}) = \frac{|J_1|}{8J_2}$
and the commensurate spiral momentum, we find
\begin{equation}
  x_{c1} = J_2 / \lambda_{c1} \approx \textrm{0.61}(J_2/J_1)^2
  \;.
\label{eq:critical1}
\end{equation}
Including non-zero 
$K_1$ or $K_2$, the critical $\lambda_{c1}$ is expected to be somewhat smaller
than the one found in Eq.~\eqref{eq:critical1}.  Since $J_2/J_1$ is
expected to be large, we have $x_{c1}>x_c = 1/16$.  

\begin{figure}
\includegraphics[width=8.0cm]{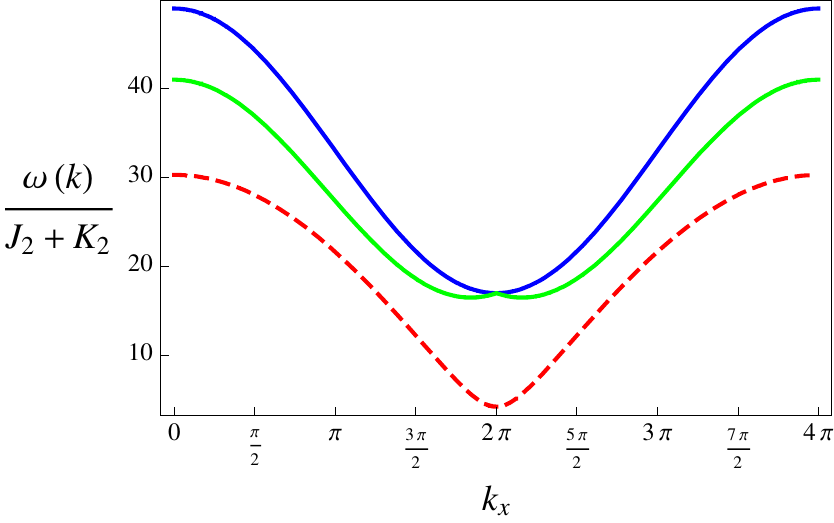}
\vskip -0.2cm
\caption{(color online).  The low-lying spectrum along $[100]$
  direction. The solid curves are calculations from the small $x$
  expansion, with $\lambda=25(J_2+K_2)$.  The blue (dark) and green
  (light) curves have $J_1+K_1= 0$ and $J_1+K_1=J_2+K_2$, respectively.
  The red (dashed) curve is a schematic spectrum close to the quantum
  critical point, where it shows a Dirac-type structure at low energy.}
\label{fig:spectrum}
\vskip -0.5cm
\end{figure}

%%%%%%%%%%%%%%%%%%%%%%%%%%%%%%%%%%%%%%%%%%%%%%%%%%%%%%%%%%%%%%%%%%%%%%%%%%%%%%%%%%%%%%%%%
\emph{Excitations in the spin orbital singlet phase} --- For small $x$,
deep in the disordered phase, one can obtain the excitation spectrum as
an expansion in the exchange.  At the leading order, we find the
lowest-lying states form a triplet, with energy
\begin{eqnarray}
  \omega({\bf k}) &=&  \lambda +  (2J_2+2K_2) \sum_{\{{\bf A}\}} \cos({\bf A} \cdot {\bf k}) 
  \nonumber \\
  & &  - | (2J_1+2K_1) \sum_{\{{\bf a}\}} \exp(i{\bf a} \cdot {\bf k})|
  \;,
\end{eqnarray}
where the ${\bf a}$ and ${\bf A}$ are summed over the 4 NN and 12 NNN
lattice vectors, respectively.  Along $[100]$ direction, the energy
minima of $\omega$ (which are also the global minima) are at $k_x =\pm 4
\arccos{[ \frac{J_1+K_1}{8(J_2+K_2)}]}$(see
Fig.~\ref{fig:spectrum}). The energy gap (for $|J_1+K_1| \leq 8(J_2+K_2)$) is
\begin{equation}
\Delta=\lambda-8(J_2+K_2)-\frac{(J_1+K_1)^2}{2(J_2+K_2)}
\; .
\label{eq:gap}
\end{equation}

\emph{Comparison with experiments} --- Perhaps the strongest indication
of proximity to a quantum critical point in \fss\, is in the neutron
scattering experiments by Krimmel et al \cite{loidl:ns,loidl:nsnmr}.  A
magnetic excitation is observed with a bandwith exceeding 20K, in
agreement with the expectations from the Curie-Weiss temperature
$\Theta_{CW}=45.1\text{K}$.  However, a much smaller gap $\Delta\approx
2\text{K}$ is observed near the minimum, the wavevector of which is
consistent with ${\bf k}=2\pi(100)$ which we have argued is most natural
for \fss.  A gap $\Delta\approx 2\text{K}$ was also observed by NMR
measurements of the $1/T_1$ relaxation of Sc nuclei \cite{loidl:nmr}.
Theoretically, we expect a double crossover between activated behavior
for $T\ll\Delta$ to linear behavior $1/T_1 \sim T$ in the quantum
critical regime $\Delta \ll T \ll T_0$, where $T_0$ is a cut-off
temperature of order $\Theta_{CW}$, and finally $1/T_1\sim
\text{const.}$ for $T\gg T_0$.  The low and high temperature limits are
clearly observed, but the quantum critical behavior is not immediately
apparent \cite{loidl:nmr}. The uniform magnetic susceptibility remains
large at low temperature despite the gap, which we take as a strong
indication of the importance of spin-orbit interaction,
i.e. Eq.~(\ref{eq:onsiteH}).  The specific heat exhibits approximate
power-law growth $C_v(T) \approx A T^{2.5}$ for 0.2K$<T<$2K, with a
linear term $C_v(T) \sim \gamma T$ below 0.2K and more complex behavior
above 2K.  Due to inversion disorder present in such spinels, we expect
the very low temperature $\gamma$ term is attributable to two-level
system defects, and the $T^{2.5}$ behavior may represent a crossover
from this to the $T^3$ magnetic contribution expected near the
QCP.  

This work suggests numerous future directions for theory and experiment.
Theoretically, the effects of fluctuations on the critical properties,
especially with $J_1 \neq 0$, warrant more detailed investigation, as do
the effects of disorder -- a relevant perturbation at the QCP.  More
theoretical studies that predict experimental signatures are also
warranted, such as possible signs of the spin-orbital singlet in the
Jahn-Teller phonon spectra.  Experimentally, it would be most exciting
to attempt to drive \fss\ past the QCP into an ordered state.  This
could perhaps be accomplished by pressure, or with a sufficiently strong
applied magnetic field.

%%%%%%%%%%%%%%%%%%%%%%%%%%%%%%%%%%%%%%%%%%%%%%%%%%%%%%%%%%%%%%%%%%%%%%%%%%%%%%%%%%%%%%%%%
\emph{Acknowledgments.} --- We would like to thank Jason Alicea for
sharing his insights.  This work was supported by the NSF grants
DMR-0804564 and PHY05-51164, and by the Packard Foundation.

\bibliography{ref} 

\end{document}